\documentstyle[pra,aps,psfig,epsfig]{revtex}
\begin{document}
\draft
\title{
\begin{flushright}
{\bf Preprint SSU-HEP-01-09\\
Samara State University}
\end{flushright}
\vspace{5mm}
\begin{center}
ONE-LOOP CORRECTIONS OF ORDER $(Z\alpha)^6m_1/m_2$, $(Z\alpha)^7$\\
TO THE MUONIUM FINE STRUCTURE\footnote{Talk presented at the XVI
International Workshop on High Energy Physics and Quantum Field Theory
(QFTHEP), Moscow, 6-12 September, 2001}
\end{center}
}
\author{R.N.Faustov\footnote{E-mail:
faustov@theory.sinp.msu.ru}}
\address{117333, Moscow, Vavilov, 40, Scientific Council "Cybernetics" RAS }
\author{A.P.Martynenko\footnote{E-mail:
mart@info.ssu.samara.ru}}
\address{443011, Samara, Pavlov, 1, Department of Theoretical Physics,
Samara State University}

\date{\today}

\maketitle

\begin{abstract}
The corrections of order $\rm {(Z\alpha)^6m_1/m_2}$ and
$\rm {(Z\alpha)^7}$ from one-loop two-photon exchange diagrams to the energy
spectra of the hydrogenic atoms are calculated with the help of the
Taylor expansion of corresponding integrands. The method of averaging
the quasipotential over the wave functions in the d-dimensional coordinate
space is formulated. The numerical values of the obtained contributions
to the fine structure of muonium, hydrogen and positronium are presented.
\end{abstract}

\immediate\write16{<<WARNING: LINEDRAW macros work with emTeX-dvivers
                    and other drivers supporting emTeX \special's
                    (dviscr, dvihplj, dvidot, dvips, dviwin, etc.) >>}

\newdimen\Lengthunit       \Lengthunit  = 1.5cm
\newcount\Nhalfperiods     \Nhalfperiods= 9
\newcount\magnitude        \magnitude = 1000

\catcode`\*=11
\newdimen\L*   \newdimen\d*   \newdimen\d**
\newdimen\dm*  \newdimen\dd*  \newdimen\dt*
\newdimen\a*   \newdimen\b*   \newdimen\c*
\newdimen\a**  \newdimen\b**
\newdimen\xL*  \newdimen\yL*
\newdimen\rx*  \newdimen\ry*
\newdimen\tmp* \newdimen\linwid*

\newcount\k*   \newcount\l*   \newcount\m*
\newcount\k**  \newcount\l**  \newcount\m**
\newcount\n*   \newcount\dn*  \newcount\r*
\newcount\N*   \newcount\*one \newcount\*two  \*one=1 \*two=2
\newcount\*ths \*ths=1000
\newcount\angle*  \newcount\q*  \newcount\q**
\newcount\angle** \angle**=0
\newcount\sc*     \sc*=0

\newtoks\cos*  \cos*={1}
\newtoks\sin*  \sin*={0}

\catcode`\[=13

\def\rotate(#1){\advance\angle**#1\angle*=\angle**
\q**=\angle*\ifnum\q**<0\q**=-\q**\fi
\ifnum\q**>360\q*=\angle*\divide\q*360\multiply\q*360\advance\angle*-\q*\fi
\ifnum\angle*<0\advance\angle*360\fi\q**=\angle*\divide\q**90\q**=\q**
\def\sgcos*{+}\def\sgsin*{+}\relax
\ifcase\q**\or
 \def\sgcos*{-}\def\sgsin*{+}\or
 \def\sgcos*{-}\def\sgsin*{-}\or
 \def\sgcos*{+}\def\sgsin*{-}\else\fi
\q*=\q**
\multiply\q*90\advance\angle*-\q*
\ifnum\angle*>45\sc*=1\angle*=-\angle*\advance\angle*90\else\sc*=0\fi
\def[##1,##2]{\ifnum\sc*=0\relax
\edef\cs*{\sgcos*.##1}\edef\sn*{\sgsin*.##2}\ifcase\q**\or
 \edef\cs*{\sgcos*.##2}\edef\sn*{\sgsin*.##1}\or
 \edef\cs*{\sgcos*.##1}\edef\sn*{\sgsin*.##2}\or
 \edef\cs*{\sgcos*.##2}\edef\sn*{\sgsin*.##1}\else\fi\else
\edef\cs*{\sgcos*.##2}\edef\sn*{\sgsin*.##1}\ifcase\q**\or
 \edef\cs*{\sgcos*.##1}\edef\sn*{\sgsin*.##2}\or
 \edef\cs*{\sgcos*.##2}\edef\sn*{\sgsin*.##1}\or
 \edef\cs*{\sgcos*.##1}\edef\sn*{\sgsin*.##2}\else\fi\fi
\cos*={\cs*}\sin*={\sn*}\global\edef\gcos*{\cs*}\global\edef\gsin*{\sn*}}\relax
\ifcase\angle*[9999,0]\or
[999,017]\or[999,034]\or[998,052]\or[997,069]\or[996,087]\or
[994,104]\or[992,121]\or[990,139]\or[987,156]\or[984,173]\or
[981,190]\or[978,207]\or[974,224]\or[970,241]\or[965,258]\or
[961,275]\or[956,292]\or[951,309]\or[945,325]\or[939,342]\or
[933,358]\or[927,374]\or[920,390]\or[913,406]\or[906,422]\or
[898,438]\or[891,453]\or[882,469]\or[874,484]\or[866,499]\or
[857,515]\or[848,529]\or[838,544]\or[829,559]\or[819,573]\or
[809,587]\or[798,601]\or[788,615]\or[777,629]\or[766,642]\or
[754,656]\or[743,669]\or[731,681]\or[719,694]\or[707,707]\or
\else[9999,0]\fi}

\catcode`\[=12

\def\GRAPH(hsize=#1)#2{\hbox to #1\Lengthunit{#2\hss}}

\def\Linewidth#1{\global\linwid*=#1\relax
\global\divide\linwid*10\global\multiply\linwid*\mag
\global\divide\linwid*100\special{em:linewidth \the\linwid*}}

\Linewidth{.4pt}
\def\sm*{\special{em:moveto}}
\def\sl*{\special{em:lineto}}
\let\moveto=\sm*
\let\lineto=\sl*
\newbox\spm*   \newbox\spl*
\setbox\spm*\hbox{\sm*}
\setbox\spl*\hbox{\sl*}

\def\mov#1(#2,#3)#4{\rlap{\L*=#1\Lengthunit
\xL*=#2\L* \yL*=#3\L*
\xL*=\xscale\xL* \yL*=\yscale\yL*
\rx* \the\cos*\xL* \tmp* \the\sin*\yL* \advance\rx*-\tmp*
\ry* \the\cos*\yL* \tmp* \the\sin*\xL* \advance\ry*\tmp*
\kern\rx*\raise\ry*\hbox{#4}}}

\def\rmov*(#1,#2)#3{\rlap{\xL*=#1\yL*=#2\relax
\rx* \the\cos*\xL* \tmp* \the\sin*\yL* \advance\rx*-\tmp*
\ry* \the\cos*\yL* \tmp* \the\sin*\xL* \advance\ry*\tmp*
\kern\rx*\raise\ry*\hbox{#3}}}

\def\lin#1(#2,#3){\rlap{\sm*\mov#1(#2,#3){\sl*}}}

\def\arr*(#1,#2,#3){\rmov*(#1\dd*,#1\dt*){\sm*
\rmov*(#2\dd*,#2\dt*){\rmov*(#3\dt*,-#3\dd*){\sl*}}\sm*
\rmov*(#2\dd*,#2\dt*){\rmov*(-#3\dt*,#3\dd*){\sl*}}}}

\def\arrow#1(#2,#3){\rlap{\lin#1(#2,#3)\mov#1(#2,#3){\relax
\d**=-.012\Lengthunit\dd*=#2\d**\dt*=#3\d**
\arr*(1,10,4)\arr*(3,8,4)\arr*(4.8,4.2,3)}}}

\def\arrlin#1(#2,#3){\rlap{\L*=#1\Lengthunit\L*=.5\L*
\lin#1(#2,#3)\rmov*(#2\L*,#3\L*){\arrow.1(#2,#3)}}}

\def\dasharrow#1(#2,#3){\rlap{{\Lengthunit=0.9\Lengthunit
\dashlin#1(#2,#3)\mov#1(#2,#3){\sm*}}\mov#1(#2,#3){\sl*
\d**=-.012\Lengthunit\dd*=#2\d**\dt*=#3\d**
\arr*(1,10,4)\arr*(3,8,4)\arr*(4.8,4.2,3)}}}

\def\clap#1{\hbox to 0pt{\hss #1\hss}}

\def\ind(#1,#2)#3{\rlap{\L*=.1\Lengthunit
\xL*=#1\L* \yL*=#2\L*
\rx* \the\cos*\xL* \tmp* \the\sin*\yL* \advance\rx*-\tmp*
\ry* \the\cos*\yL* \tmp* \the\sin*\xL* \advance\ry*\tmp*
\kern\rx*\raise\ry*\hbox{\lower2pt\clap{$#3$}}}}

\def\sh*(#1,#2)#3{\rlap{\dm*=\the\n*\d**
\xL*=\xscale\dm* \yL*=\yscale\dm* \xL*=#1\xL* \yL*=#2\yL*
\rx* \the\cos*\xL* \tmp* \the\sin*\yL* \advance\rx*-\tmp*
\ry* \the\cos*\yL* \tmp* \the\sin*\xL* \advance\ry*\tmp*
\kern\rx*\raise\ry*\hbox{#3}}}

\def\calcnum*#1(#2,#3){\a*=1000sp\b*=1000sp\a*=#2\a*\b*=#3\b*
\ifdim\a*<0pt\a*-\a*\fi\ifdim\b*<0pt\b*-\b*\fi
\ifdim\a*>\b*\c*=.96\a*\advance\c*.4\b*
\else\c*=.96\b*\advance\c*.4\a*\fi
\k*\a*\multiply\k*\k*\l*\b*\multiply\l*\l*
\m*\k*\advance\m*\l*\n*\c*\r*\n*\multiply\n*\n*
\dn*\m*\advance\dn*-\n*\divide\dn*2\divide\dn*\r*
\advance\r*\dn*
\c*=\the\Nhalfperiods5sp\c*=#1\c*\ifdim\c*<0pt\c*-\c*\fi
\multiply\c*\r*\N*\c*\divide\N*10000}

\def\dashlin#1(#2,#3){\rlap{\calcnum*#1(#2,#3)\relax
\d**=#1\Lengthunit\ifdim\d**<0pt\d**-\d**\fi
\divide\N*2\multiply\N*2\advance\N*\*one
\divide\d**\N*\sm*\n*\*one\sh*(#2,#3){\sl*}\loop
\advance\n*\*one\sh*(#2,#3){\sm*}\advance\n*\*one
\sh*(#2,#3){\sl*}\ifnum\n*<\N*\repeat}}

\def\dashdotlin#1(#2,#3){\rlap{\calcnum*#1(#2,#3)\relax
\d**=#1\Lengthunit\ifdim\d**<0pt\d**-\d**\fi
\divide\N*2\multiply\N*2\advance\N*1\multiply\N*2\relax
\divide\d**\N*\sm*\n*\*two\sh*(#2,#3){\sl*}\loop
\advance\n*\*one\sh*(#2,#3){\kern-1.48pt\lower.5pt\hbox{\rm.}}\relax
\advance\n*\*one\sh*(#2,#3){\sm*}\advance\n*\*two
\sh*(#2,#3){\sl*}\ifnum\n*<\N*\repeat}}

\def\shl*(#1,#2)#3{\kern#1#3\lower#2#3\hbox{\unhcopy\spl*}}

\def\trianglin#1(#2,#3){\rlap{\toks0={#2}\toks1={#3}\calcnum*#1(#2,#3)\relax
\dd*=.57\Lengthunit\dd*=#1\dd*\divide\dd*\N*
\divide\dd*\*ths \multiply\dd*\magnitude
\d**=#1\Lengthunit\ifdim\d**<0pt\d**-\d**\fi
\multiply\N*2\divide\d**\N*\sm*\n*\*one\loop
\shl**{\dd*}\dd*-\dd*\advance\n*2\relax
\ifnum\n*<\N*\repeat\n*\N*\shl**{0pt}}}

\def\wavelin#1(#2,#3){\rlap{\toks0={#2}\toks1={#3}\calcnum*#1(#2,#3)\relax
\dd*=.23\Lengthunit\dd*=#1\dd*\divide\dd*\N*
\divide\dd*\*ths \multiply\dd*\magnitude
\d**=#1\Lengthunit\ifdim\d**<0pt\d**-\d**\fi
\multiply\N*4\divide\d**\N*\sm*\n*\*one\loop
\shl**{\dd*}\dt*=1.3\dd*\advance\n*\*one
\shl**{\dt*}\advance\n*\*one
\shl**{\dd*}\advance\n*\*two
\dd*-\dd*\ifnum\n*<\N*\repeat\n*\N*\shl**{0pt}}}

\def\w*lin(#1,#2){\rlap{\toks0={#1}\toks1={#2}\d**=\Lengthunit\dd*=-.12\d**
\divide\dd*\*ths \multiply\dd*\magnitude
\N*8\divide\d**\N*\sm*\n*\*one\loop
\shl**{\dd*}\dt*=1.3\dd*\advance\n*\*one
\shl**{\dt*}\advance\n*\*one
\shl**{\dd*}\advance\n*\*one
\shl**{0pt}\dd*-\dd*\advance\n*1\ifnum\n*<\N*\repeat}}

\def\l*arc(#1,#2)[#3][#4]{\rlap{\toks0={#1}\toks1={#2}\d**=\Lengthunit
\dd*=#3.037\d**\dd*=#4\dd*\dt*=#3.049\d**\dt*=#4\dt*\ifdim\d**>10mm\relax
\d**=.25\d**\n*\*one\shl**{-\dd*}\n*\*two\shl**{-\dt*}\n*3\relax
\shl**{-\dd*}\n*4\relax\shl**{0pt}\else
\ifdim\d**>5mm\d**=.5\d**\n*\*one\shl**{-\dt*}\n*\*two
\shl**{0pt}\else\n*\*one\shl**{0pt}\fi\fi}}

\def\d*arc(#1,#2)[#3][#4]{\rlap{\toks0={#1}\toks1={#2}\d**=\Lengthunit
\dd*=#3.037\d**\dd*=#4\dd*\d**=.25\d**\sm*\n*\*one\shl**{-\dd*}\relax
\n*3\relax\sh*(#1,#2){\xL*=\xscale\dd*\yL*=\yscale\dd*
\kern#2\xL*\lower#1\yL*\hbox{\sm*}}\n*4\relax\shl**{0pt}}}

\def\shl**#1{\c*=\the\n*\d**\d*=#1\relax
\a*=\the\toks0\c*\b*=\the\toks1\d*\advance\a*-\b*
\b*=\the\toks1\c*\d*=\the\toks0\d*\advance\b*\d*
\a*=\xscale\a*\b*=\yscale\b*
\rx* \the\cos*\a* \tmp* \the\sin*\b* \advance\rx*-\tmp*
\ry* \the\cos*\b* \tmp* \the\sin*\a* \advance\ry*\tmp*
\raise\ry*\rlap{\kern\rx*\unhcopy\spl*}}

\def\wlin*#1(#2,#3)[#4]{\rlap{\toks0={#2}\toks1={#3}\relax
\c*=#1\l*\c*\c*=.01\Lengthunit\m*\c*\divide\l*\m*
\c*=\the\Nhalfperiods5sp\multiply\c*\l*\N*\c*\divide\N*\*ths
\divide\N*2\multiply\N*2\advance\N*\*one
\dd*=.002\Lengthunit\dd*=#4\dd*\multiply\dd*\l*\divide\dd*\N*
\divide\dd*\*ths \multiply\dd*\magnitude
\d**=#1\multiply\N*4\divide\d**\N*\sm*\n*\*one\loop
\shl**{\dd*}\dt*=1.3\dd*\advance\n*\*one
\shl**{\dt*}\advance\n*\*one
\shl**{\dd*}\advance\n*\*two
\dd*-\dd*\ifnum\n*<\N*\repeat\n*\N*\shl**{0pt}}}

\def\wavebox#1{\setbox0\hbox{#1}\relax
\a*=\wd0\advance\a*14pt\b*=\ht0\advance\b*\dp0\advance\b*14pt\relax
\hbox{\kern9pt\relax
\rmov*(0pt,\ht0){\rmov*(-7pt,7pt){\wlin*\a*(1,0)[+]\wlin*\b*(0,-1)[-]}}\relax
\rmov*(\wd0,-\dp0){\rmov*(7pt,-7pt){\wlin*\a*(-1,0)[+]\wlin*\b*(0,1)[-]}}\relax
\box0\kern9pt}}

\def\rectangle#1(#2,#3){\relax
\lin#1(#2,0)\lin#1(0,#3)\mov#1(0,#3){\lin#1(#2,0)}\mov#1(#2,0){\lin#1(0,#3)}}

\def\dashrectangle#1(#2,#3){\dashlin#1(#2,0)\dashlin#1(0,#3)\relax
\mov#1(0,#3){\dashlin#1(#2,0)}\mov#1(#2,0){\dashlin#1(0,#3)}}

\def\waverectangle#1(#2,#3){\L*=#1\Lengthunit\a*=#2\L*\b*=#3\L*
\ifdim\a*<0pt\a*-\a*\def\x*{-1}\else\def\x*{1}\fi
\ifdim\b*<0pt\b*-\b*\def\y*{-1}\else\def\y*{1}\fi
\wlin*\a*(\x*,0)[-]\wlin*\b*(0,\y*)[+]\relax
\mov#1(0,#3){\wlin*\a*(\x*,0)[+]}\mov#1(#2,0){\wlin*\b*(0,\y*)[-]}}

\def\calcparab*{\ifnum\n*>\m*\k*\N*\advance\k*-\n*\else\k*\n*\fi
\a*=\the\k* sp\a*=10\a*\b*\dm*\advance\b*-\a*\k*\b*
\a*=\the\*ths\b*\divide\a*\l*\multiply\a*\k*
\divide\a*\l*\k*\*ths\r*\a*\advance\k*-\r*\dt*=\the\k*\L*}

\def\arcto#1(#2,#3)[#4]{\rlap{\toks0={#2}\toks1={#3}\calcnum*#1(#2,#3)\relax
\dm*=135sp\dm*=#1\dm*\d**=#1\Lengthunit\ifdim\dm*<0pt\dm*-\dm*\fi
\multiply\dm*\r*\a*=.3\dm*\a*=#4\a*\ifdim\a*<0pt\a*-\a*\fi
\advance\dm*\a*\N*\dm*\divide\N*10000\relax
\divide\N*2\multiply\N*2\advance\N*\*one
\L*=-.25\d**\L*=#4\L*\divide\d**\N*\divide\L*\*ths
\m*\N*\divide\m*2\dm*=\the\m*5sp\l*\dm*\sm*\n*\*one\loop
\calcparab*\shl**{-\dt*}\advance\n*1\ifnum\n*<\N*\repeat}}

\def\arrarcto#1(#2,#3)[#4]{\L*=#1\Lengthunit\L*=.54\L*
\arcto#1(#2,#3)[#4]\rmov*(#2\L*,#3\L*){\d*=.457\L*\d*=#4\d*\d**-\d*
\rmov*(#3\d**,#2\d*){\arrow.02(#2,#3)}}}

\def\dasharcto#1(#2,#3)[#4]{\rlap{\toks0={#2}\toks1={#3}\relax
\calcnum*#1(#2,#3)\dm*=\the\N*5sp\a*=.3\dm*\a*=#4\a*\ifdim\a*<0pt\a*-\a*\fi
\advance\dm*\a*\N*\dm*
\divide\N*20\multiply\N*2\advance\N*1\d**=#1\Lengthunit
\L*=-.25\d**\L*=#4\L*\divide\d**\N*\divide\L*\*ths
\m*\N*\divide\m*2\dm*=\the\m*5sp\l*\dm*
\sm*\n*\*one\loop\calcparab*
\shl**{-\dt*}\advance\n*1\ifnum\n*>\N*\else\calcparab*
\sh*(#2,#3){\xL*=#3\dt* \yL*=#2\dt*
\rx* \the\cos*\xL* \tmp* \the\sin*\yL* \advance\rx*\tmp*
\ry* \the\cos*\yL* \tmp* \the\sin*\xL* \advance\ry*-\tmp*
\kern\rx*\lower\ry*\hbox{\sm*}}\fi
\advance\n*1\ifnum\n*<\N*\repeat}}

\def\*shl*#1{\c*=\the\n*\d**\advance\c*#1\a**\d*\dt*\advance\d*#1\b**
\a*=\the\toks0\c*\b*=\the\toks1\d*\advance\a*-\b*
\b*=\the\toks1\c*\d*=\the\toks0\d*\advance\b*\d*
\rx* \the\cos*\a* \tmp* \the\sin*\b* \advance\rx*-\tmp*
\ry* \the\cos*\b* \tmp* \the\sin*\a* \advance\ry*\tmp*
\raise\ry*\rlap{\kern\rx*\unhcopy\spl*}}

\def\calcnormal*#1{\b**=10000sp\a**\b**\k*\n*\advance\k*-\m*
\multiply\a**\k*\divide\a**\m*\a**=#1\a**\ifdim\a**<0pt\a**-\a**\fi
\ifdim\a**>\b**\d*=.96\a**\advance\d*.4\b**
\else\d*=.96\b**\advance\d*.4\a**\fi
\d*=.01\d*\r*\d*\divide\a**\r*\divide\b**\r*
\ifnum\k*<0\a**-\a**\fi\d*=#1\d*\ifdim\d*<0pt\b**-\b**\fi
\k*\a**\a**=\the\k*\dd*\k*\b**\b**=\the\k*\dd*}

\def\wavearcto#1(#2,#3)[#4]{\rlap{\toks0={#2}\toks1={#3}\relax
\calcnum*#1(#2,#3)\c*=\the\N*5sp\a*=.4\c*\a*=#4\a*\ifdim\a*<0pt\a*-\a*\fi
\advance\c*\a*\N*\c*\divide\N*20\multiply\N*2\advance\N*-1\multiply\N*4\relax
\d**=#1\Lengthunit\dd*=.012\d**
\divide\dd*\*ths \multiply\dd*\magnitude
\ifdim\d**<0pt\d**-\d**\fi\L*=.25\d**
\divide\d**\N*\divide\dd*\N*\L*=#4\L*\divide\L*\*ths
\m*\N*\divide\m*2\dm*=\the\m*0sp\l*\dm*
\sm*\n*\*one\loop\calcnormal*{#4}\calcparab*
\*shl*{1}\advance\n*\*one\calcparab*
\*shl*{1.3}\advance\n*\*one\calcparab*
\*shl*{1}\advance\n*2\dd*-\dd*\ifnum\n*<\N*\repeat\n*\N*\shl**{0pt}}}

\def\triangarcto#1(#2,#3)[#4]{\rlap{\toks0={#2}\toks1={#3}\relax
\calcnum*#1(#2,#3)\c*=\the\N*5sp\a*=.4\c*\a*=#4\a*\ifdim\a*<0pt\a*-\a*\fi
\advance\c*\a*\N*\c*\divide\N*20\multiply\N*2\advance\N*-1\multiply\N*2\relax
\d**=#1\Lengthunit\dd*=.012\d**
\divide\dd*\*ths \multiply\dd*\magnitude
\ifdim\d**<0pt\d**-\d**\fi\L*=.25\d**
\divide\d**\N*\divide\dd*\N*\L*=#4\L*\divide\L*\*ths
\m*\N*\divide\m*2\dm*=\the\m*0sp\l*\dm*
\sm*\n*\*one\loop\calcnormal*{#4}\calcparab*
\*shl*{1}\advance\n*2\dd*-\dd*\ifnum\n*<\N*\repeat\n*\N*\shl**{0pt}}}

\def\hr*#1{\L*=\xscale\Lengthunit\ifnum
\angle**=0\clap{\vrule width#1\L* height.1pt}\else
\L*=#1\L*\L*=.5\L*\rmov*(-\L*,0pt){\sm*}\rmov*(\L*,0pt){\sl*}\fi}

\def\shade#1[#2]{\rlap{\Lengthunit=#1\Lengthunit
\special{em:linewidth .001pt}\relax
\mov(0,#2.05){\hr*{.994}}\mov(0,#2.1){\hr*{.980}}\relax
\mov(0,#2.15){\hr*{.953}}\mov(0,#2.2){\hr*{.916}}\relax
\mov(0,#2.25){\hr*{.867}}\mov(0,#2.3){\hr*{.798}}\relax
\mov(0,#2.35){\hr*{.715}}\mov(0,#2.4){\hr*{.603}}\relax
\mov(0,#2.45){\hr*{.435}}\special{em:linewidth \the\linwid*}}}

\def\dshade#1[#2]{\rlap{\special{em:linewidth .001pt}\relax
\Lengthunit=#1\Lengthunit\if#2-\def\t*{+}\else\def\t*{-}\fi
\mov(0,\t*.025){\relax
\mov(0,#2.05){\hr*{.995}}\mov(0,#2.1){\hr*{.988}}\relax
\mov(0,#2.15){\hr*{.969}}\mov(0,#2.2){\hr*{.937}}\relax
\mov(0,#2.25){\hr*{.893}}\mov(0,#2.3){\hr*{.836}}\relax
\mov(0,#2.35){\hr*{.760}}\mov(0,#2.4){\hr*{.662}}\relax
\mov(0,#2.45){\hr*{.531}}\mov(0,#2.5){\hr*{.320}}\relax
\special{em:linewidth \the\linwid*}}}}

\def\vdot{\rlap{\kern-1.9pt\lower1.8pt\hbox{$\scriptstyle\bullet$}}}
\def\vtimes{\rlap{\kern-3pt\lower1.8pt\hbox{$\scriptstyle\times$}}}
\def\vDot{\rlap{\kern-2.3pt\lower2.7pt\hbox{$\bullet$}}}
\def\vTimes{\rlap{\kern-3.6pt\lower2.4pt\hbox{$\times$}}}

\def\arc(#1)[#2,#3]{{\k*=#2\l*=#3\m*=\l*
\advance\m*-6\ifnum\k*>\l*\relax\else
{\rotate(#2)\mov(#1,0){\sm*}}\loop
\ifnum\k*<\m*\advance\k*5{\rotate(\k*)\mov(#1,0){\sl*}}\repeat
{\rotate(#3)\mov(#1,0){\sl*}}\fi}}

\def\dasharc(#1)[#2,#3]{{\k**=#2\n*=#3\advance\n*-1\advance\n*-\k**
\L*=1000sp\L*#1\L* \multiply\L*\n* \multiply\L*\Nhalfperiods
\divide\L*57\N*\L* \divide\N*2000\ifnum\N*=0\N*1\fi
\r*\n*  \divide\r*\N* \ifnum\r*<2\r*2\fi
\m**\r* \divide\m**2 \l**\r* \advance\l**-\m** \N*\n* \divide\N*\r*
\k**\r* \multiply\k**\N* \dn*\n* \advance\dn*-\k** \divide\dn*2\advance\dn*\*one
\r*\l** \divide\r*2\advance\dn*\r* \advance\N*-2\k**#2\relax
\ifnum\l**<6{\rotate(#2)\mov(#1,0){\sm*}}\advance\k**\dn*
{\rotate(\k**)\mov(#1,0){\sl*}}\advance\k**\m**
{\rotate(\k**)\mov(#1,0){\sm*}}\loop
\advance\k**\l**{\rotate(\k**)\mov(#1,0){\sl*}}\advance\k**\m**
{\rotate(\k**)\mov(#1,0){\sm*}}\advance\N*-1\ifnum\N*>0\repeat
{\rotate(#3)\mov(#1,0){\sl*}}\else\advance\k**\dn*
\arc(#1)[#2,\k**]\loop\advance\k**\m** \r*\k**
\advance\k**\l** {\arc(#1)[\r*,\k**]}\relax
\advance\N*-1\ifnum\N*>0\repeat
\advance\k**\m**\arc(#1)[\k**,#3]\fi}}

\def\triangarc#1(#2)[#3,#4]{{\k**=#3\n*=#4\advance\n*-\k**
\L*=1000sp\L*#2\L* \multiply\L*\n* \multiply\L*\Nhalfperiods
\divide\L*57\N*\L* \divide\N*1000\ifnum\N*=0\N*1\fi
\d**=#2\Lengthunit \d*\d** \divide\d*57\multiply\d*\n*
\r*\n*  \divide\r*\N* \ifnum\r*<2\r*2\fi
\m**\r* \divide\m**2 \l**\r* \advance\l**-\m** \N*\n* \divide\N*\r*
\dt*\d* \divide\dt*\N* \dt*.5\dt* \dt*#1\dt*
\divide\dt*1000\multiply\dt*\magnitude
\k**\r* \multiply\k**\N* \dn*\n* \advance\dn*-\k** \divide\dn*2\relax
\r*\l** \divide\r*2\advance\dn*\r* \advance\N*-1\k**#3\relax
{\rotate(#3)\mov(#2,0){\sm*}}\advance\k**\dn*
{\rotate(\k**)\mov(#2,0){\sl*}}\advance\k**-\m**\advance\l**\m**\loop\dt*-\dt*
\d*\d** \advance\d*\dt*
\advance\k**\l**{\rotate(\k**)\rmov*(\d*,0pt){\sl*}}%
\advance\N*-1\ifnum\N*>0\repeat\advance\k**\m**
{\rotate(\k**)\mov(#2,0){\sl*}}{\rotate(#4)\mov(#2,0){\sl*}}}}

\def\wavearc#1(#2)[#3,#4]{{\k**=#3\n*=#4\advance\n*-\k**
\L*=4000sp\L*#2\L* \multiply\L*\n* \multiply\L*\Nhalfperiods
\divide\L*57\N*\L* \divide\N*1000\ifnum\N*=0\N*1\fi
\d**=#2\Lengthunit \d*\d** \divide\d*57\multiply\d*\n*
\r*\n*  \divide\r*\N* \ifnum\r*=0\r*1\fi
\m**\r* \divide\m**2 \l**\r* \advance\l**-\m** \N*\n* \divide\N*\r*
\dt*\d* \divide\dt*\N* \dt*.7\dt* \dt*#1\dt*
\divide\dt*1000\multiply\dt*\magnitude
\k**\r* \multiply\k**\N* \dn*\n* \advance\dn*-\k** \divide\dn*2\relax
\divide\N*4\advance\N*-1\k**#3\relax
{\rotate(#3)\mov(#2,0){\sm*}}\advance\k**\dn*
{\rotate(\k**)\mov(#2,0){\sl*}}\advance\k**-\m**\advance\l**\m**\loop\dt*-\dt*
\d*\d** \advance\d*\dt* \dd*\d** \advance\dd*1.3\dt*
\advance\k**\r*{\rotate(\k**)\rmov*(\d*,0pt){\sl*}}\relax
\advance\k**\r*{\rotate(\k**)\rmov*(\dd*,0pt){\sl*}}\relax
\advance\k**\r*{\rotate(\k**)\rmov*(\d*,0pt){\sl*}}\relax
\advance\k**\r*
\advance\N*-1\ifnum\N*>0\repeat\advance\k**\m**
{\rotate(\k**)\mov(#2,0){\sl*}}{\rotate(#4)\mov(#2,0){\sl*}}}}

\def\gmov*#1(#2,#3)#4{\rlap{\L*=#1\Lengthunit
\xL*=#2\L* \yL*=#3\L*
\rx* \gcos*\xL* \tmp* \gsin*\yL* \advance\rx*-\tmp*
\ry* \gcos*\yL* \tmp* \gsin*\xL* \advance\ry*\tmp*
\rx*=\xscale\rx* \ry*=\yscale\ry*
\xL* \the\cos*\rx* \tmp* \the\sin*\ry* \advance\xL*-\tmp*
\yL* \the\cos*\ry* \tmp* \the\sin*\rx* \advance\yL*\tmp*
\kern\xL*\raise\yL*\hbox{#4}}}

\def\rgmov*(#1,#2)#3{\rlap{\xL*#1\yL*#2\relax
\rx* \gcos*\xL* \tmp* \gsin*\yL* \advance\rx*-\tmp*
\ry* \gcos*\yL* \tmp* \gsin*\xL* \advance\ry*\tmp*
\rx*=\xscale\rx* \ry*=\yscale\ry*
\xL* \the\cos*\rx* \tmp* \the\sin*\ry* \advance\xL*-\tmp*
\yL* \the\cos*\ry* \tmp* \the\sin*\rx* \advance\yL*\tmp*
\kern\xL*\raise\yL*\hbox{#3}}}

\def\Earc(#1)[#2,#3][#4,#5]{{\k*=#2\l*=#3\m*=\l*
\advance\m*-6\ifnum\k*>\l*\relax\else\def\xscale{#4}\def\yscale{#5}\relax
{\angle**0\rotate(#2)}\gmov*(#1,0){\sm*}\loop
\ifnum\k*<\m*\advance\k*5\relax
{\angle**0\rotate(\k*)}\gmov*(#1,0){\sl*}\repeat
{\angle**0\rotate(#3)}\gmov*(#1,0){\sl*}\relax
\def\xscale{1}\def\yscale{1}\fi}}

\def\dashEarc(#1)[#2,#3][#4,#5]{{\k**=#2\n*=#3\advance\n*-1\advance\n*-\k**
\L*=1000sp\L*#1\L* \multiply\L*\n* \multiply\L*\Nhalfperiods
\divide\L*57\N*\L* \divide\N*2000\ifnum\N*=0\N*1\fi
\r*\n*  \divide\r*\N* \ifnum\r*<2\r*2\fi
\m**\r* \divide\m**2 \l**\r* \advance\l**-\m** \N*\n* \divide\N*\r*
\k**\r*\multiply\k**\N* \dn*\n* \advance\dn*-\k** \divide\dn*2\advance\dn*\*one
\r*\l** \divide\r*2\advance\dn*\r* \advance\N*-2\k**#2\relax
\ifnum\l**<6\def\xscale{#4}\def\yscale{#5}\relax
{\angle**0\rotate(#2)}\gmov*(#1,0){\sm*}\advance\k**\dn*
{\angle**0\rotate(\k**)}\gmov*(#1,0){\sl*}\advance\k**\m**
{\angle**0\rotate(\k**)}\gmov*(#1,0){\sm*}\loop
\advance\k**\l**{\angle**0\rotate(\k**)}\gmov*(#1,0){\sl*}\advance\k**\m**
{\angle**0\rotate(\k**)}\gmov*(#1,0){\sm*}\advance\N*-1\ifnum\N*>0\repeat
{\angle**0\rotate(#3)}\gmov*(#1,0){\sl*}\def\xscale{1}\def\yscale{1}\else
\advance\k**\dn* \Earc(#1)[#2,\k**][#4,#5]\loop\advance\k**\m** \r*\k**
\advance\k**\l** {\Earc(#1)[\r*,\k**][#4,#5]}\relax
\advance\N*-1\ifnum\N*>0\repeat
\advance\k**\m**\Earc(#1)[\k**,#3][#4,#5]\fi}}

\def\triangEarc#1(#2)[#3,#4][#5,#6]{{\k**=#3\n*=#4\advance\n*-\k**
\L*=1000sp\L*#2\L* \multiply\L*\n* \multiply\L*\Nhalfperiods
\divide\L*57\N*\L* \divide\N*1000\ifnum\N*=0\N*1\fi
\d**=#2\Lengthunit \d*\d** \divide\d*57\multiply\d*\n*
\r*\n*  \divide\r*\N* \ifnum\r*<2\r*2\fi
\m**\r* \divide\m**2 \l**\r* \advance\l**-\m** \N*\n* \divide\N*\r*
\dt*\d* \divide\dt*\N* \dt*.5\dt* \dt*#1\dt*
\divide\dt*1000\multiply\dt*\magnitude
\k**\r* \multiply\k**\N* \dn*\n* \advance\dn*-\k** \divide\dn*2\relax
\r*\l** \divide\r*2\advance\dn*\r* \advance\N*-1\k**#3\relax
\def\xscale{#5}\def\yscale{#6}\relax
{\angle**0\rotate(#3)}\gmov*(#2,0){\sm*}\advance\k**\dn*
{\angle**0\rotate(\k**)}\gmov*(#2,0){\sl*}\advance\k**-\m**
\advance\l**\m**\loop\dt*-\dt* \d*\d** \advance\d*\dt*
\advance\k**\l**{\angle**0\rotate(\k**)}\rgmov*(\d*,0pt){\sl*}\relax
\advance\N*-1\ifnum\N*>0\repeat\advance\k**\m**
{\angle**0\rotate(\k**)}\gmov*(#2,0){\sl*}\relax
{\angle**0\rotate(#4)}\gmov*(#2,0){\sl*}\def\xscale{1}\def\yscale{1}}}

\def\waveEarc#1(#2)[#3,#4][#5,#6]{{\k**=#3\n*=#4\advance\n*-\k**
\L*=4000sp\L*#2\L* \multiply\L*\n* \multiply\L*\Nhalfperiods
\divide\L*57\N*\L* \divide\N*1000\ifnum\N*=0\N*1\fi
\d**=#2\Lengthunit \d*\d** \divide\d*57\multiply\d*\n*
\r*\n*  \divide\r*\N* \ifnum\r*=0\r*1\fi
\m**\r* \divide\m**2 \l**\r* \advance\l**-\m** \N*\n* \divide\N*\r*
\dt*\d* \divide\dt*\N* \dt*.7\dt* \dt*#1\dt*
\divide\dt*1000\multiply\dt*\magnitude
\k**\r* \multiply\k**\N* \dn*\n* \advance\dn*-\k** \divide\dn*2\relax
\divide\N*4\advance\N*-1\k**#3\def\xscale{#5}\def\yscale{#6}\relax
{\angle**0\rotate(#3)}\gmov*(#2,0){\sm*}\advance\k**\dn*
{\angle**0\rotate(\k**)}\gmov*(#2,0){\sl*}\advance\k**-\m**
\advance\l**\m**\loop\dt*-\dt*
\d*\d** \advance\d*\dt* \dd*\d** \advance\dd*1.3\dt*
\advance\k**\r*{\angle**0\rotate(\k**)}\rgmov*(\d*,0pt){\sl*}\relax
\advance\k**\r*{\angle**0\rotate(\k**)}\rgmov*(\dd*,0pt){\sl*}\relax
\advance\k**\r*{\angle**0\rotate(\k**)}\rgmov*(\d*,0pt){\sl*}\relax
\advance\k**\r*
\advance\N*-1\ifnum\N*>0\repeat\advance\k**\m**
{\angle**0\rotate(\k**)}\gmov*(#2,0){\sl*}\relax
{\angle**0\rotate(#4)}\gmov*(#2,0){\sl*}\def\xscale{1}\def\yscale{1}}}

\newcount\CatcodeOfAtSign
\CatcodeOfAtSign=\the\catcode`\@
\catcode`\@=11
\def\@arc#1[#2][#3]{\rlap{\Lengthunit=#1\Lengthunit
\sm*\l*arc(#2.1914,#3.0381)[#2][#3]\relax
\mov(#2.1914,#3.0381){\l*arc(#2.1622,#3.1084)[#2][#3]}\relax
\mov(#2.3536,#3.1465){\l*arc(#2.1084,#3.1622)[#2][#3]}\relax
\mov(#2.4619,#3.3086){\l*arc(#2.0381,#3.1914)[#2][#3]}}}

\def\dash@arc#1[#2][#3]{\rlap{\Lengthunit=#1\Lengthunit
\d*arc(#2.1914,#3.0381)[#2][#3]\relax
\mov(#2.1914,#3.0381){\d*arc(#2.1622,#3.1084)[#2][#3]}\relax
\mov(#2.3536,#3.1465){\d*arc(#2.1084,#3.1622)[#2][#3]}\relax
\mov(#2.4619,#3.3086){\d*arc(#2.0381,#3.1914)[#2][#3]}}}

\def\wave@arc#1[#2][#3]{\rlap{\Lengthunit=#1\Lengthunit
\w*lin(#2.1914,#3.0381)\relax
\mov(#2.1914,#3.0381){\w*lin(#2.1622,#3.1084)}\relax
\mov(#2.3536,#3.1465){\w*lin(#2.1084,#3.1622)}\relax
\mov(#2.4619,#3.3086){\w*lin(#2.0381,#3.1914)}}}

\def\bezier#1(#2,#3)(#4,#5)(#6,#7){\N*#1\l*\N* \advance\l*\*one
\d* #4\Lengthunit \advance\d* -#2\Lengthunit \multiply\d* \*two
\b* #6\Lengthunit \advance\b* -#2\Lengthunit
\advance\b*-\d* \divide\b*\N*
\d** #5\Lengthunit \advance\d** -#3\Lengthunit \multiply\d** \*two
\b** #7\Lengthunit \advance\b** -#3\Lengthunit
\advance\b** -\d** \divide\b**\N*
\mov(#2,#3){\sm*{\loop\ifnum\m*<\l*
\a*\m*\b* \advance\a*\d* \divide\a*\N* \multiply\a*\m*
\a**\m*\b** \advance\a**\d** \divide\a**\N* \multiply\a**\m*
\rmov*(\a*,\a**){\unhcopy\spl*}\advance\m*\*one\repeat}}}

\catcode`\*=12

\newcount\n@ast
\def\n@ast@#1{\n@ast0\relax\get@ast@#1\end}
\def\get@ast@#1{\ifx#1\end\let\next\relax\else
\ifx#1*\advance\n@ast1\fi\let\next\get@ast@\fi\next}

\newif\if@up \newif\if@dwn
\def\up@down@#1{\@upfalse\@dwnfalse
\if#1u\@uptrue\fi\if#1U\@uptrue\fi\if#1+\@uptrue\fi
\if#1d\@dwntrue\fi\if#1D\@dwntrue\fi\if#1-\@dwntrue\fi}

\def\halfcirc#1(#2)[#3]{{\Lengthunit=#2\Lengthunit\up@down@{#3}\relax
\if@up\mov(0,.5){\@arc[-][-]\@arc[+][-]}\fi
\if@dwn\mov(0,-.5){\@arc[-][+]\@arc[+][+]}\fi
\def\lft{\mov(0,.5){\@arc[-][-]}\mov(0,-.5){\@arc[-][+]}}\relax
\def\rght{\mov(0,.5){\@arc[+][-]}\mov(0,-.5){\@arc[+][+]}}\relax
\if#3l\lft\fi\if#3L\lft\fi\if#3r\rght\fi\if#3R\rght\fi
\n@ast@{#1}\relax
\ifnum\n@ast>0\if@up\shade[+]\fi\if@dwn\shade[-]\fi\fi
\ifnum\n@ast>1\if@up\dshade[+]\fi\if@dwn\dshade[-]\fi\fi}}

\def\halfdashcirc(#1)[#2]{{\Lengthunit=#1\Lengthunit\up@down@{#2}\relax
\if@up\mov(0,.5){\dash@arc[-][-]\dash@arc[+][-]}\fi
\if@dwn\mov(0,-.5){\dash@arc[-][+]\dash@arc[+][+]}\fi
\def\lft{\mov(0,.5){\dash@arc[-][-]}\mov(0,-.5){\dash@arc[-][+]}}\relax
\def\rght{\mov(0,.5){\dash@arc[+][-]}\mov(0,-.5){\dash@arc[+][+]}}\relax
\if#2l\lft\fi\if#2L\lft\fi\if#2r\rght\fi\if#2R\rght\fi}}

\def\halfwavecirc(#1)[#2]{{\Lengthunit=#1\Lengthunit\up@down@{#2}\relax
\if@up\mov(0,.5){\wave@arc[-][-]\wave@arc[+][-]}\fi
\if@dwn\mov(0,-.5){\wave@arc[-][+]\wave@arc[+][+]}\fi
\def\lft{\mov(0,.5){\wave@arc[-][-]}\mov(0,-.5){\wave@arc[-][+]}}\relax
\def\rght{\mov(0,.5){\wave@arc[+][-]}\mov(0,-.5){\wave@arc[+][+]}}\relax
\if#2l\lft\fi\if#2L\lft\fi\if#2r\rght\fi\if#2R\rght\fi}}

\catcode`\*=11

\def\Circle#1(#2){\halfcirc#1(#2)[u]\halfcirc#1(#2)[d]\n@ast@{#1}\relax
\ifnum\n@ast>0\L*=\xscale\Lengthunit
\ifnum\angle**=0\clap{\vrule width#2\L* height.1pt}\else
\L*=#2\L*\L*=.5\L*\special{em:linewidth .001pt}\relax
\rmov*(-\L*,0pt){\sm*}\rmov*(\L*,0pt){\sl*}\relax
\special{em:linewidth \the\linwid*}\fi\fi}

\catcode`\*=12

\def\wavecirc(#1){\halfwavecirc(#1)[u]\halfwavecirc(#1)[d]}

\def\dashcirc(#1){\halfdashcirc(#1)[u]\halfdashcirc(#1)[d]}

\def\xscale{1}
\def\yscale{1}

\def\Ellipse#1(#2)[#3,#4]{\def\xscale{#3}\def\yscale{#4}\relax
\Circle#1(#2)\def\xscale{1}\def\yscale{1}}

\def\dashEllipse(#1)[#2,#3]{\def\xscale{#2}\def\yscale{#3}\relax
\dashcirc(#1)\def\xscale{1}\def\yscale{1}}

\def\waveEllipse(#1)[#2,#3]{\def\xscale{#2}\def\yscale{#3}\relax
\wavecirc(#1)\def\xscale{1}\def\yscale{1}}

\def\halfEllipse#1(#2)[#3][#4,#5]{\def\xscale{#4}\def\yscale{#5}\relax
\halfcirc#1(#2)[#3]\def\xscale{1}\def\yscale{1}}

\def\halfdashEllipse(#1)[#2][#3,#4]{\def\xscale{#3}\def\yscale{#4}\relax
\halfdashcirc(#1)[#2]\def\xscale{1}\def\yscale{1}}

\def\halfwaveEllipse(#1)[#2][#3,#4]{\def\xscale{#3}\def\yscale{#4}\relax
\halfwavecirc(#1)[#2]\def\xscale{1}\def\yscale{1}}

\catcode`\@=\the\CatcodeOfAtSign

\section{Introduction}

The study of perturbative effects of higher order in $\alpha$
($\alpha$ is the fine structure constant) in the energy spectra of the
positronium and muonium is of great practical importance for
testing the bound state theory in quantum electrodynamics. In the past few
years considerable progress has been made towards that goal \cite{MT}.
First of all it is determined by an increase of the accuracy in the measurement
of the fine and hyperfine structure in these leptonic systems.
Thus, the experimental error for the muonium ground state hyperfine splitting
was reduced by a factor of three \cite{Liu}:
\begin{equation}
{\rm \Delta \nu^{exp}_{HFS} (Mu)= 4~463~302~765(53)~Hz}
\end{equation}
Even larger increase of the precision was reached in the measurement of
the interval ${\rm 1S\div 2S}$ in the muonium on the basis of the
Doppler-free two-photon spectroscopy \cite{Meyer}:
\begin{equation}
\rm {\Delta \nu^{exp}_{Mu}(2^3S_1\div 1^3S_1)= 2~455~528~941.0(9.8)~MHz.}
\end{equation}

The value of the same interval in positronium obtained some years ago
is the following \cite{Fee}:
\begin{equation}
{\rm \Delta \nu^{exp}_{Ps}(2^3S_1\div 1^3S_1)=1~233~607~216.4(3.2)~MHz.}
\end{equation}

One of the most precise experiment in the spectroscopy of simple
atomic systems is the measurement of the hydrogen gross structure
${\rm 1S\div 2S}$ \cite{Udem}:
\begin{equation}
\rm {\Delta \nu_H(2S-1S)=2~466~061~413~187.34(84) ~kHz}
\end{equation}

On the other hand this progress results from the development of
the computer methods for the calculation of Feynman amplitudes
and the appearance of nonrelativistic quantum electrodynamics
(NRQED) \cite{CL,PL,Nio} for the calculation of the bound state
energy spectra at the nonrelativistic scale. Due to
performed calculations there were obtained some new contributions
to the energy levels of hydrogen-like systems: corrections of
order ${\rm (Z\alpha)^6m_1/m_2}$ to the muonium fine structure
\cite{EGS,EG,VMS,VS,AY,MF}, corrections of order $\rm{m\alpha^6}$
to the positronium fine and hyperfine structure
\cite{CMY,A1,A2,PK,AB}, logarithmic contributions ${\rm
O(m\alpha^7 ln^2\alpha)}$ to the positronium spectrum
\cite{MY,Penin} and also the logarithmic corrections of order
${\rm O(\alpha^7\ln\alpha)}$ to the muonium and the positronium
hyperfine splittings \cite{K93,MY1}. Calculation of loop integrals
which represent the corrections to the Coulomb potential by the
perturbation theory is complicated problem because of the presence of
the essentially different energy scales determined the behaviour
of the integrands. The following four scales can be introduced in
terms of the loop energy ${\rm p^0}$ and the loop three momentum
${\rm {\bf p}}$ \cite{DY,BYG}:
\begin{eqnarray}
{\rm hard~momentum~region: \rm {p^0\sim\mu, |{\bf p}|\sim\mu,}}
\end{eqnarray}
\begin{eqnarray}
{\rm potential~region: \rm {p^0\sim\mu\alpha^2, |{\bf p}|\sim\mu\alpha,}}
\end{eqnarray}
\begin{eqnarray}
{\rm soft~momentum~region: \rm {p^0\sim\mu\alpha, |{\bf p}|\sim\mu\alpha,}}
\end{eqnarray}
\begin{eqnarray}
{\rm ultrasoft~momentum~region: \rm {p^0\sim\mu\alpha^2, |{\bf p}|\sim\mu\alpha^2,}}
\end{eqnarray}
where the mass parameter $\mu$ is determined by the masses of the particles
${\rm m_1}$, ${\rm m_2}$. There are two approaches to the calculation
of the definite order contribution on $\alpha$. the first one is related with
the explicit extraction of the small parameter in the corresponding QED Feynman
amplitudes which than can be expanded in Taylor series in one of the
energy regions (5)-(8) \cite{MF,CS,Tkachov,BS}. The second approach
realized in NRQED consists in the formulation of such
procedure already at the Lagrangian level \cite{CL,PL,Nio}. In present
work which is the sequel of \cite{MF} we study the contributions of order
${\rm (Z\alpha)^6}$ and ${\rm (Z\alpha)^7}$ to the spectra of muonium
and positronium from one-loop Feynman amplitudes using the first approach.
The method of dimensional regularization and Taylor series allows to
construct the interaction operator of the particles in the momentum and
the coordinate representations. Numerically the order factors are equal to
${\rm (Z\alpha)^6m_e^2/m_\mu}$=0.0902 MHz and
${\rm m_e\alpha^7}$=0.136 MHz so the calculation of the mentioned
contributions is very actual especially for their confrontation with
experimental data (1) and (4).

\section{Contributions of order $\rm {(Z\alpha)^6m_1/m_2}$ to the
muonium fine structure.}

Simple counting of the $Z\alpha$ powers in the two-photon exchange diagrams
in Fig.1 shows, that they can contribute in the order
${\rm (Z\alpha)^2\cdot(Z\alpha)^3=(Z\alpha)^5}$. Here the first factor
is connected with exchanged photons and the second one is determined by
the wave functions of the bound state. At the same time the
availability of the relative motion momentum of the particles in the
initial state ${\rm {\bf p}}$ (${\rm |{\bf p}|\sim\mu\alpha}$)
and in the final state ${\rm{\bf q}}$ (${\rm |{\bf q}|\sim\mu\alpha}$)
in these $2\gamma$ amplitudes lead to the appearance of the higher order
contributions on $\alpha$. First of all we formulate here the approach
to the calculation of definite order corrections and then calculate
the contributions containing additional degree ${\rm Z\alpha}$ and the ratio
of the electron to muon masses ${\rm m_1/m_2=0.004836}$. This approach
can be used also for the positronium spectrum if we don't employ some
simplifications in the $2\gamma$ Feynman amplitudes connected with small value
${\rm m_1/m_2}$. The general structure of the integrals describing
the quasipotential remains unchanged in this case.

Our calculations are based on a local quasipotential equation of the
Schroedinger type \cite{MF}:
\begin{eqnarray}
{\rm \left(\frac{b^2}{2\mu_R}-\frac{{\bf p}^2}{2\mu_R}\right)\psi_M({\bf p})=
\int\frac{d{\bf q}}{(2\pi)^3}V({\bf p},{\bf q},M)\psi_M({\bf q}),}
\end{eqnarray}
where ${\rm b^2=E_1^2-m_1^2}$=${\rm E_2^2-m_2^2}$, ${\rm \mu_R=E_1E_2/M}$
is the relativistic reduced mass, ${\rm M=E_1+E_2}$ is the mass of the bound
state, ${\rm \psi_M(\bf p)}$ is the quasipotential wave function.
For the initial approximation of the quasipotential ${\rm V({\bf p},
{\bf q},M)}$ for the bound system we take the ordinary Coulomb potential.

To construct the interaction operator of the particles corresponding
to $2\gamma$ amplitudes we used also the projection operators
${\rm \hat{\pi}_{S=0}}$ and ${\rm \hat{\pi}_{S=1}}$ on the states with total
spin S=0 and 1 in the system ${\rm (e^-\mu^+)}$:
\begin{equation}
{\rm \hat{\pi}_{S=0}=\frac{\hat{P}+M}{2\sqrt{2}M}\gamma_5,~~~
\hat{\pi}_{S=1}=\frac{\hat{P}+M}{2\sqrt{2}M}\hat{\epsilon},}
\end{equation}
where ${\rm \epsilon^\mu}$ is the polarization vector for the state
${\rm ^3S_1}$, ${\rm\hat\epsilon=\epsilon^\mu\gamma_\mu}$.

\begin{figure*}[t!]
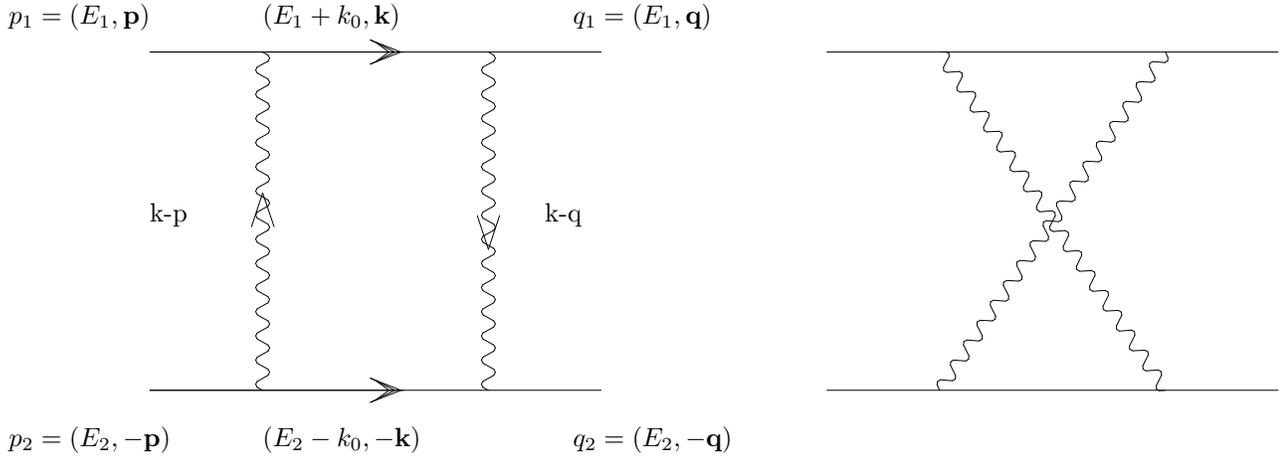

\magnitude=2000
\GRAPH(hsize=15){
\mov(1,0){\lin(4,0)}%
\mov(-0.25,-0.5){$p_2=(E_2,-{\bf p})$}%
\mov(-0.25,3.25){$p_1=(E_1,{\bf p})$}%
\mov(4.75,-0.5){$q_2=(E_2,-{\bf q})$}%
\mov(4.75,3.25){$q_1=(E_1,{\bf q})$}%
\mov(1,1.5){k-p}%
\mov(4.5,1.5){k-q}%
\mov(2,3.25){$(E_1+k_0,{\bf k})$}%
\mov(2,-0.5){$(E_2-k_0,-{\bf k})$}%
\mov(1,3){\lin(4,0)}%
\mov(7,0){\lin(4,0)}
\mov(7,3){\lin(4,0)}
\mov(1,0){\arrow(2.25,0)}%
\mov(1,3){\arrow(2.25,0)}
\mov(2,0){\wavelin(0,3)}%
\mov(2,1.75){\lin(-0.1,-0.3)}%
\mov(2,1.75){\lin(0.1,-0.3)}%
\mov(4,1.25){\lin(-0.1,0.3)}%
\mov(4,1.25){\lin(0.1,0.3)}%
\mov(4,0){\wavelin(0,3)}%
\mov(8,0){\wavelin(2,3)}%
\mov(8,3){\wavelin(2,-3)}%
}
\vspace{2mm}
\caption{
Two-photon exchange Feynman diagrams. ${\rm P=p_1+p_2}$ is the total
momentum of the two-particle bound state, p,q are the relative
four-momenta of particles in the initial and final states.}
\end{figure*}

The contribution of diagram (a) Fig.1 takes the form (we consider further
only triplet states ${\rm ^3S_1}$):
\begin{eqnarray}
\rm {V_{2\gamma}^{(a)}({\bf p},{\bf
q})=\frac{i(Z\alpha)^2}{\pi^2}\int\frac{f_1(k,m_1,m_2)d^4k}
{(k-p)^2(k-q)^2D_1(k)D_2(-k)},}
\end{eqnarray}
\begin{displaymath}
\rm {f_1(k,m_1,m_2)=m_2(4m_1+2k^0)+2m_1k^0-2{k^0}^2+\frac{2}{3}{\bf k}^2.}
\end{displaymath}
The denominator of the electron propagator ${\rm D_1(k)}$=
${\rm k^2+2m_1k^0+b^2}$.
But corresponding muon denominator can be simplified taking into account
the small parameter ${\rm m_1/m_2}$ and neglecting by the muon kinetic energy
in the intermediate state: ${\rm D_2(-k)\approx 2m_2(-k^0+i\epsilon)}$.
The crossed two-photon diagram (b) Fig.1 gives the similar contribution:
\begin{eqnarray}
\rm {V_{2\gamma}^{(b)}({\bf p},{\bf
q})=\frac{i(Z\alpha)^2}{\pi^2}\int\frac{f_2(k,m_1,m_2)d^4k}
{(k-p)^2(k-q)^2D_1(k)D_2(-k+p+q)},}
\end{eqnarray}
\begin{displaymath}
\rm
{f_2(k,m_1,m_2)=m_2(4m_1+2k^0)+2m_1k^0+6{k^0}^2+\frac{10}{3}({\bf
p}{\bf k}+ {\bf q}{\bf k}-{\bf k}^2).}
\end{displaymath}
Direct two-photon amplitude (a) Fig.1 contains the contribution of lower
order on $\alpha$ ($\sim\alpha^3$) which is cancelled by the iteration
part of the quasipotential ${\rm V_{1\gamma}\times G^f\times V_{1\gamma}}$
\cite{MF}. To extract such term we transform the product of the electron
and muon denominators as follows \cite{BYG}:
\begin{eqnarray}
{\rm \frac{1}{D_1(k)D_2(-k)}=\frac{-2\pi i\delta(k^0)}{-2E({\bf k}^2-b^2)}-
\frac{1}{2E}\left[\frac{1}{(k^0+i0)D_1(k)}+\frac{1}{(-k^0+i0)D_2(-k)}\right].}
\end{eqnarray}
Omitting the first term in the right part of (12) corresponding to free
two-particle propagator we can present the sum of potentials
${\rm V_{2\gamma}^{(a)}}$+${\rm V_{2\gamma}^{(b)}}$ in the form:
\begin{eqnarray}
{\rm V_{2\gamma}({\bf p},{\bf q})=V_{2\gamma}^{(a)}({\bf p},{\bf q})+
V_{2\gamma}^{(b)}({\bf p},{\bf q})=
\frac{i(Z\alpha)^2}{6m_2\pi^2}\int\frac{d^4 k}{(k-p)^2(k-q)^2}\times}
\end{eqnarray}
\begin{displaymath}
\rm {\times\left[\frac{12m_1^2+18m_1k^0+24{k^0}^2-12{\bf k}^2+10{\bf k}({\bf p}+{\bf q})}
{(k^0+i0)D_1(k)}-\frac{6m_1+3k^0}{(-k^0+i0)^2}\right].}
\end{displaymath}
Consider first of all the necessary order contribution of (14) in the
potential region (6). In this case the integration over ${\rm k^0}$
is determined by the residue in the pole of the electron propagator,
which can be presented as Taylor expansion:
\begin{eqnarray}
\rm {\frac{1}{{k^0}^2-{\bf k}^2+2m_1k^0+b^2}=\sum_{n=0}^\infty\frac{(-{k^0}^2)^n}
{(-{\bf k}^2+2m_1k^0+b^2)^{n+1}}.}
\end{eqnarray}
The expansion of photon propagators in the potential region looks as
follows:
\begin{eqnarray}
\rm {\frac{1}{{k^0}^2-({\bf k}-{\bf
p})^2}=\sum_{n=0}^\infty\frac{(-{k^0}^2)^n} {[-({\bf k}-{\bf
p})^2]^{n+1}}.}
\end{eqnarray}
Multiplying the expansions (15) and (16) we observe the next structure
of (14):
\begin{eqnarray}
\rm {V^{pot}_{2\gamma}({\bf p},{\bf q})\sim\int d^4k\left[\frac{1}
{-({\bf k}-{\bf p})^2}-
\frac{{k^0}^2}{({\bf k}-{\bf p})^4}-\frac{{k^0}^4}{({\bf k}-{\bf p})^6}-
\ldots\right]\times}
\end{eqnarray}
\begin{displaymath}
{\rm \times\left[\frac{1}{-({\bf k}-{\bf q})^2}-\frac{{k^0}^2}{({\bf k}
-{\bf q})^4}-
\frac{{k^0}^4}{({\bf k}-{\bf q})^6}-\ldots\right]\frac{1}{(k^0+i0)}\times}
\end{displaymath}
\begin{displaymath}
{\rm \times\left[\frac{1}{-{\bf k}^2+2m_1k^0+b^2}-\frac{{k^0}^2}{(-{\bf k}^2
+2m_1k^0+b^2)^2}-\frac{{k^0}^4}{(-{\bf k}^2+2m_1k^0+b^2)^3}-\ldots\right].}
\end{displaymath}
The integral (17) contains corrections of different order on $\alpha$
which is determined by the corresponding factors after integration over
${\rm k^0}$ in the pole of electron propagator.
Existing contribution of order ${\rm (Z\alpha)^6}$ in Eq. (17) is determined
by the following typical integral:
\begin{eqnarray}
{\rm J({\bf p},{\bf q})=\int\frac{d{\bf k}}{(2\pi)^3}\frac{({\bf k}^2)^\alpha}
{({\bf k}^2-2{\bf k}{\bf p}+{\bf p}^2)^\beta({\bf k}^2-2{\bf k}{\bf q}+
{\bf q}^2)^\gamma},}
\end{eqnarray}
where the degrees $\alpha, \beta, \gamma$ can be different, but the sum
${\rm 2\alpha+3-2\beta-2\gamma=1}$.
The contribution to the energy spectrum of this system can be obtained
after averaging ${\rm J({\bf p},{\bf q})}$ over Coulomb wave functions.
The characteristic integrals (18) are divergent just as in ultraviolet
region so also in infrared one. The reason of these divergences lies
in the used Taylor expansions of the integral function in Eq. (14) which
we made to extract the necessary order contribution on $\alpha$.
The initial integral (14) is finite so all divergences in the intermediate
expressions of the type (18) must be cancelled in the sum. There are two
ways for the calculation of the average value ${\rm <J({\bf p},{\bf q})>}$:

1. The integration (18) over ${\rm {\bf k}}$ can be done by means of the
dimensional regularization. Extracting the finite part of the result we
can construct the Fourier transform and averaging it on Coulomb wave
functions in the r-space.

2. Carrying out the Fourier transform of Eq. (18) directly we can then
average it in coordinate space.

We used both approaches for the calculation of the contributions
${\rm (Z\alpha)^6m_1/m_2}$ in the muonium energy spectrum. The transition
to the d-dimensional space allows to regularize the expression of the
quasipotential both in the relativistic region of the intermediate momenta
and in nonrelativistic one. The results of the integrations (18) for
different functions are presented in the Table I.

The region of the soft momenta (7) gives the necessary order contributions
to the muonium S-states. In this case the value of the integral (14)
is determined by the residue in the photon poles. The expansion of the
electron propagator in the region (7) takes the form:
\begin{eqnarray}
\rm {\frac{1}{{k^0}^2-{\bf k}^2+2m_1k^0+b^2}=\sum_{n=0}^\infty\frac{({k^0}^2-{\bf k}^2+b^2)^n}
{(2m_1k^0)^{n+1}}.}
\end{eqnarray}
Using it we have the following standard integrals in Eq. (14):
\begin{eqnarray}
\rm {I_{soft}({\bf p},{\bf q})\sim\sum_{n=0}^\infty\int\frac{d^4k}{(k-p)^2(k-q)^2}
\frac{1}{(k^0+i0)}\frac{({k^0}^2-{\bf k}^2+b^2)^n}{(2m_1k^0)^{n+1}}.}
\end{eqnarray}
Since the function in the numerator of Eq. (14) contains the terms
of order 1, $\alpha$, $\alpha^2$, the sum (20) gives the contributions of order
${\rm O(\alpha^6)}$ when n=1, 2, 3. The subsequent calculation of such
integrals was carried out as in the potential region (6).

\section{Matrix elements of the operator $r^{-\nu}$ in d-dimensional
coordinate space.}

In many cases the averaging of the two-photon interaction quasipotential
on Coulomb wave functions can be conveniently done in coordinate
representation. Indeed the calculation of the integrals ${\rm J({\bf p},{\bf q})}$
shows that the matrix elements ${\rm <J({\bf p},{\bf q})>}$ in three dimensional
momentum space are divergent. The reason for these divergences is the same:
Taylor expansions of the integrands. We can implement again the
dimensional regularization as for the integration on loop momentum
${\rm {\bf k}}$. But then there is need to know the two-particle wave
function for the Coulomb potential in d-dimensional momentum space
\cite{Popov,SA}. Other approach for the calculation of the average value
${\rm <V_{2\gamma}>}$ is connected with the construction of Fourier
transform for the potential (14) in d-dimensional coordinate representation
and subsequent calculation of the matrix elements on Coulomb wave
functions in r-space. The Fourier transform of the power potential
${\rm V(p)=1/p^m}$ is determined by the following expression \cite{PR}:
\begin{equation}
\rm {V(r)=\frac{1}{(2\pi)^{d-1}}\prod_{k=1}^{d-3}\frac{\Gamma(\frac{k+1}{2})
\Gamma(1/2)}{\Gamma(\frac{k+2}{2})}\int_0^\pi\int_0^\infty \sin^{d-2}\theta
e^{ipr \cos\theta}d\theta p^{d-1-m} dp
=\frac{\Gamma(\frac{d-m}{2})}{2^m\pi^{d/2}r^{d-m}\Gamma(\frac{m}{2})}.}
\end{equation}
So in three dimensional coordinate space d=3 we have the degree
potentials which lead to the divergent matrix elements for the case of
S-states. To extract such singular terms of the kind
$1/\epsilon$ (${\rm d=3+2\epsilon}$) we formulate auxiliary Coulomb
task in d-dimensional coordinate space. Let consider the ordinary
Coulomb potential in d-dimensional space \cite{Popov,SA}:
\begin{eqnarray}
{\rm H=\frac{{\bf p}^2}{2}+V_C(r)=\frac{p_r^2}{2}+\frac{1}{2 r^2}
\left[{\bf L}^2+\frac{(d-1)(d-3)}{4}\right]- \frac{a}{r},~~a=\mu
Z\alpha,~~r=\sqrt{\sum_{i=1}^dx_i^2},}
\end{eqnarray}
where ${\rm p_r}$ is radial momentum operator, $\mu$ is nonrelativistic
reduced mass. The proper functions
of the angular part ${\rm {\bf L}^2}$ of the Laplace operator $\Delta$
in d-dimensional space are the homogeneous harmonic polynomials of the degree
l on sphere ${\rm S^{d-1}}$ with proper values ${\rm \lambda=l(l+d-2)}$ \cite{NV}.
The exact solution of the d-dimensional Coulomb problem in the case
of discrete spectrum was obtained in Ref. \cite{SA}. In particular it
was shown that Coulomb energy levels are the following:
\begin{eqnarray}
{\rm \epsilon_n=-\frac{1}{2\left(n+\frac{d-3}{2}\right)^2},}
\end{eqnarray}
where n=1, 2, 3, ... is the principal quantum number (see Appendix A).
The Coulomb Green function in d-dimensional case is important for the
calculations of the corrections in the higher orders of the perturbation
theory. The expression for it was obtained in Ref. \cite{PP}. The
reqqurence relation for the matrix elements of the power operators
$r^{-\nu}$ are very important for finding the average values of the
quasipotential ${\rm V_{2\gamma}}$. To obtain this relation in d-dimensional
space consider the operator of the radial momentum ${\rm p_r}$:
\begin{equation}
\rm {p_r=\frac{1}{2}\left\{\frac{r_i}{r},p_i\right\}=\frac{r_i}{r}p_i-\frac{i(d-1)}{2r}=
-i\left(\frac{\partial}{\partial r}+\frac{d-1}{2r}\right).}
\end{equation}
It satisfies to the following commutation relations:
\begin{equation}
\rm {[r,p_r]=i\cdot I,~~~\left[\frac{1}{r^\nu},p_r\right]=\frac{-i\nu}{r^{\nu+1}}.}
\end{equation}
From the expressions (24), (25) it appears that:
\begin{equation}
\rm {\left[H,\frac{1}{r^{\nu-1}}\right]=\frac{\nu-1}{2}\left(-\frac{\nu}{r^{\nu+1}}+
\frac{2i}{r^\nu}p_r\right),~~~
[H, ip_r]=\frac{\Delta_0}{r^3}-\frac{a}{r^2},~~\Delta_0={\bf L}^2+\frac{(d-1)(d-3)}{4}.}
\end{equation}
Another important relation can be obtained from (24)-(26) \cite{Bohm}:
\begin{eqnarray}
\rm {\left[\frac{1}{r^\nu}\left[r,H\right],H\right]+\frac{\nu}{2}
\left[\frac{1}{r^{\nu+1}},H\right]=
\frac{2\nu}{r^{\nu+1}}H-\frac{\nu+1}{r^{\nu+3}}\left[{\bf L}^2+\frac{(d-1)
(d-3)}{4}\right]+}
\end{eqnarray}
\begin{displaymath}
{\rm +\frac{(2\nu+1)a}{r^{\nu+2}}+\frac{\nu(\nu+1)
(\nu+2)}{4}\frac{1}{r^{\nu+3}}.}
\end{displaymath}
So the reqqurence relation for the matrix elements ${\rm \frac{1}{r^\nu}}$
can be derived after averaging (27) on the wave functions ${\rm \psi_{nlm}}$
satisfying the following properties:
\begin{eqnarray}
\rm {H\psi_{nlm}=E_n\psi_{nlm},~~
E_n=a^2\epsilon_n=-\frac{a^2}{2(n+\frac{d-3}{2})^2}, ~~{\bf
L}^2\psi_{nlm}=l(l+d-2)\psi_{nlm}.}
\end{eqnarray}
The matrix element of the left part (27) on $\psi_{nlm}$ is equal to 0.
So we obtain the necessary reqqurence relation in the form:
\begin{equation}
\rm {0=2\nu E_n<\frac{1}{r^{\nu+1}}>-(\nu+1)\left[l(l+d-2)+\frac{(d-1)(d-3)}{4}
\right]<\frac{1}{r^{\nu+3}}>+}
\end{equation}
\begin{displaymath}
\rm {+a(2\nu+1)
<\frac{1}{r^{\nu+2}}>+\frac{\nu(\nu+1)(\nu+2)}{4}<\frac{1}{r^{\nu+3}}>.}
\end{displaymath}
In the case of S-states l=0 and the obtained reqqurence relation (29)
can be rewritten as follows:
\begin{equation}
{\rm
<\frac{1}{r^{\nu+3}}>=-\frac{8E_n\nu}{(\nu+1)[\nu(\nu+2)-(d-1)(d-3)]}
<\frac{1}{r^{\nu+1}}>-}
\end{equation}
\begin{displaymath}
{\rm -\frac{4a(2\nu+1)}{(\nu+1)[\nu(\nu+2)-(d-1)(d-3)]}<\frac{1}{r^{\nu+2}}>.}
\end{displaymath}
This formula allows to express the matrix elements of high negative powers
of the radius ${\rm \frac{1}{r^\nu}}$ which are singular in d=3 space
through the matrix elements of lower powers r extracting explicitly
the singular factors ${\rm \frac{1}{\epsilon}}$ $\rm {(d=3+2\epsilon)}$.
As an example for the application of (30) we consider the calculation
of the typical integral in (14) using two independent approaches:
\begin{eqnarray}
{\rm I({\bf p},{\bf q})=\int\frac{d{\bf
k}}{(2\pi)^d}\frac{(4\pi\alpha)^2({\bf k}^2-b^2)} {({\bf k}-{\bf
p})^2({\bf k}-{\bf q})^22\mu},}
\end{eqnarray}
where the factors ${\rm (4\pi\alpha)^2}$ and ${\rm 1/2\mu}$ were
introduced to simplify the intermediate expressions. Indeed the Coulomb
wave function satisfies to the Eq. (9) with Coulomb potential \cite{MF1}.
Using it we can transform (31) to the sum of the different matrix
elements:
\begin{eqnarray}
{\rm <I({\bf p},{\bf q})>=<\left(\frac{{\bf k}^2-b^2}{2\mu}\right)^3>=
<({\bf k}^6+3{\bf k}^4W^2+3{\bf k}^2W^4+W^6)\frac{1}{8\mu^3}>, ~~~W^2=-b^2.}
\end{eqnarray}
Some of them can be calculated directly in d=3 space:
\begin{eqnarray}
{\rm <{\bf k}^2>=<2\mu(E_n-V_C)>=\frac{(\mu\alpha)^2}{n^2}, ~~~<W^6>=\frac{(\mu\alpha)^6}{n^6},}
\end{eqnarray}
\begin{eqnarray}
{\rm <{\bf k}^4>=<(E_n-V_C)(E_n-V_C)4\mu^2>=(\mu\alpha)^4\left[-\frac{3}{n^4}+\frac{8}{n^3}\right].}
\end{eqnarray}
But the different situation arises when we average ${\rm {\bf k}^6}$:
\begin{eqnarray}
{\rm <\frac{{\bf k}^6}{8\mu^3}>=<(E_n-V_C)\frac{{\bf k}^2}{2\mu}(E_n-V_C)>=
<E_n^2(E_n-V_C)>-}
\end{eqnarray}
\begin{displaymath}
{\rm -2<E_n(E_n-V_C)V_C>-<\frac{W^2}{2\mu}V_C^2>-<V_C^3>+
<\frac{\alpha^2}{2r^4}>.}
\end{displaymath}
There are two divergent matrix elements in three dimensional space:
${\rm \sim<\frac{1}{r^3}>}$ and ${\rm <\frac{1}{r^4}>}$. Using our
reqqurence relation (30) we observe that the singular parts of (35) at
are cancelled in the limit ${\rm d\rightarrow 3}$. We have the following
finite result for it:
\begin{eqnarray}
{\rm <\frac{{\bf k}^6}{8\mu^3}>=\mu^3\alpha^6\left[\frac{5}{8n^6}-
\frac{7}{3n^5}-\frac{8}{3n^3}\right].}
\end{eqnarray}
Summing the contributions (33)-(34), (36) in the expression (32)
we find the value of the initial integral:
\begin{eqnarray}
{\rm <I({\bf p},{\bf q})>=\mu^3\alpha^6
\left[\frac{2}{3n^5}-\frac{8}{3n^3}\right].}
\end{eqnarray}
Consider the alternative method for the calculation of (31). We can carry
out direct integration in (31) by means of dimensional regularization.
Then going to the d=3 we obtain the following contribution to the
quasipotential in momentum space:
\begin{eqnarray}
{\rm I({\bf p},{\bf q})=\frac{(4\pi\alpha)^2}{2\mu}\left[\frac{{\bf p}^2-b^2}
{8t}-\frac{t}{16}\right], {\bf t}=({\bf p}-{\bf q}).}
\end{eqnarray}
The Fourier transform of (38) allows to derive the corresponding potential
in the coordinate representation:
\begin{eqnarray}
{\rm I(r)=\frac{\alpha^2}{2\mu}\left[({\bf p}^2-b^2)\frac{1}{r^2}+
\frac{1}{r^4}\right].}
\end{eqnarray}
Average values of the terms in this relation can be found using the equation
for the wave function and reqqurence relation (30):
\begin{eqnarray}
{\rm <\frac{({\bf p}^2-b^2)}{2\mu}\frac{\alpha^2}{r^2}>=<\frac{\alpha^3}{r^3}>=
\mu^3\alpha^6\left[\frac{2}{\epsilon n^3}-\frac{6}{n^4}-\frac{6}{n^3}\right],}
\end{eqnarray}
\begin{eqnarray}
{\rm <\frac{\alpha^2}{2\mu r^4}>=\frac{\mu^3\alpha^6}{2}
\left[-\frac{4}{\epsilon n^3}+\frac{20}{3n^3}+\frac{12}{n^4}
+\frac{4}{3n^5}\right],}
\end{eqnarray}
where the contributions ${\rm \sim 1/\epsilon}$ in two last divergent
matrix elements are distinguished only by the sign. As a result the sum
of (40) and (41) is finite and coincides with (37). Both methods for the
calculation of the integrals similar to ${\rm I({\bf p},{\bf q})}$
were applied in this work to obtain the contribution of order
${\rm (Z\alpha)^6m_1/m_2}$ in the muonium spectrum. Total sum of all
contributions in the potential region (6) and soft region (7) to the energy
levels is equal
\begin{equation}
{\rm <V_{2\gamma}>=\frac{m_1(Z\alpha)^6}{m_2}\left[-\frac{37}{9n^3}-
\frac{1}{6n^4}+\frac{7}{9n^5}\right].}
\end{equation}

\begin{table}[p]
\caption{Results of the integrations (18),~
${\rm {\bf t}=({\bf p}-{\bf q})}$.}
\bigskip
\begin{tabular}{|c|c|c|c|}     \hline
$\alpha$ & 0 & 1 & 2 \\   \hline
         &   &   &    \\
$\beta=1, \gamma=1$ & $\frac{1}{8t}$ & $\frac{{\bf p}^2+{\bf q}^2-{\bf t}^2}{16t}$ & --- \\
         &   &   &    \\   \hline
         &   &   &    \\
$\beta=1, \gamma=2$ & --- & $\frac{{\bf p}^2-{\bf q}^2+{\bf t}^2}{16t^3}$ & $\frac{{\bf p}^4-3({\bf q}^2-{\bf t}^2)^2+2{\bf p}^2({\bf q}^2+{\bf t}^2)}{32t^3} $  \\
        &    &   &    \\   \hline
        &    &   &    \\
$\beta=1, \gamma=3$ & --- & --- & $\frac{3{\bf p}^4+3({\bf q}^2-{\bf t}^2)^2+{\bf p}^2(-6{\bf q}^2+2{\bf t}^2)}{64t^5} $ \\
        &    &   &    \\   \hline
        &    &   &    \\
$\beta=2, \gamma=2$ & --- & --- & $\frac{{\bf t}^4+2{\bf t}^2({\bf p}^2+{\bf q}^2)-3({\bf p}^2-{\bf q}^2)^2}{32t^5}$ \\
        &     &   &   \\   \hline
\end{tabular}
\end{table}

\section{Corrections of order $(Z\alpha)^7$ to S energy levels.}

As was pointed out above the main contribution of two-photon amplitudes
to the energy spectrum is of order ${\rm O(\alpha^5)}$. These amplitudes
give also the contribution of order ${\rm (Z\alpha)^7}$ in the hard
momentum region. To obtain such corrections to the S-levels of the bound
state we average ${\rm 2\gamma}$ amplitudes over the particle spin. As a
result the ${\rm 2\gamma}$ quasipotentials (a) and (b) Fig.1 can be
presented in the form:
\begin{eqnarray}
{\rm V_{2\gamma}^{(a)}({\bf p},{\bf
q})=\frac{i(Z\alpha)^2}{\pi^2}\int\frac{g_1(k,m_1,m_2)d^4k}
{(k-p)^2(k-q)^2D_1(k)D_2(-k)},}
\end{eqnarray}
\begin{displaymath}
\rm {g_1(k,m_1,m_2)=16m_1m_2-8m_1k^0+8m_2k^0-4W^2\left(\frac{m_1}{m_2}+
\frac{m_2}{m_1}\right)+}
\end{displaymath}
\begin{displaymath}
\rm {+8k^0W^2\left(\frac{1}{m_1}-\frac{1}{m_2}\right)
-16{k^0}^2+8{\bf k}^2,}
\end{displaymath}
\begin{eqnarray}
{\rm V_{2\gamma}^{(b)}({\bf p},{\bf
q})=\frac{i(Z\alpha)^2}{\pi^2}\int\frac{g_2(k,m_1,m_2)d^4k}
{(k-p)^2(k-q)^2D_1(k)D_2(-k+p+q)},}
\end{eqnarray}
\begin{displaymath}
\rm {g_2(k,m_1,m_2)=16m_1m_2+8m_1k^0+8m_2k^0-4W^2\left(\frac{m_1}{m_2}+\frac{m_2}{m_1}\right)-}
\end{displaymath}
\begin{displaymath}
\rm {-8k^0W^2\left(\frac{1}{m_1}-\frac{1}{m_2}\right)
+16{k^0}^2-8{\bf k}^2+8{\bf k}{\bf p}+8{\bf k}{\bf q},}
\end{displaymath}
\begin{equation}
\rm {D_{1,2}(k)=k^2+2E_{1,2}k^0-W^2.}
\end{equation}
Taking into account that ${\rm k^0\sim\mu}$, ${\rm |{\bf k}|\sim\mu}$
the expressions of the electron and photon propagators are the following:
\begin{equation}
\rm {\frac{1}{D_{1,2}(k)}=\sum_{n=0}^\infty\frac{(W^2)^n}{(k^2+2E_{1,2}k^0)^{n+1}},}
\end{equation}
\begin{eqnarray}
\rm {\frac{1}{(k-p)^2}\approx\frac{1}{k^2}+\frac{{\bf p}^2-2{\bf p}{\bf k}}
{(k^2)^2}+\frac{4({\bf k}{\bf p})^2}{(k^2)^3},}
\end{eqnarray}
where we kept only the terms of the necessary order in the last expression.
Substituting (45) and (47) to (43), (44) we revealed that the hard part
of the quasipotential ${\rm V_{2\gamma}}$ is determined by the set of the
characteristic integrals:
\begin{eqnarray}
{\rm K({\bf p},{\bf q})=\int\frac{d^4k}{16i\pi^2}\frac{S(k)}{(k^2)^\alpha(k^2+2k^0E_1)^\beta
(k^2\pm 2k^0E_2)^\gamma},}
\end{eqnarray}

\begin{table}
\caption{The results of some basic integrals (48):
$\int\frac{d^4k}{16i\pi^2}\frac{S(k)}{(k^2)^\alpha(k^2+2k^0E_1)^\beta
(k^2-2k^0E_2)^\gamma}$}
\bigskip
\begin{tabular}{|c|c|c|c|c|}     \hline
S(k) & 1 & $k^0$ & ${\bf k}^2$ & ${k^0}^2$  \\   \hline
     &   &       &            &              \\
$\alpha=2, \beta=1, \gamma=1$ & $\frac{E_1E_2-E_1^2-E_2^2}{96E_1^3E_2^3}$ &
$\frac{E_1-E_2}{32E_1^2E_2^2}$ & $\frac{1}{32E_1E_2}$ & --- \\
     &   &       &            &           \\   \hline
     &   &       &            &            \\
$\alpha=2, \beta=2, \gamma=1$ &$\frac{2E_1^2E_2-3E_1E_2^2-E_1^3+4E_2^3}{960E_1^5E_2^4}$ &
$\frac{2E_1E_2-E_1^2-3E_2^2}{192E_1^4E_2^3}$ & --- &$\frac{E_1^2-E_2^2-E_1E_2}{128E_1^3E_2^2(E_1+E_2)}$  \\
     &   &       &            &          \\    \hline
     &   &       &            &          \\
$\alpha=3, \beta=1, \gamma=1$ &---&$\frac{(E_1^2+E_2^2)(E_1-E_2)}{192E_1^4E_2^4}$ &
$\frac{E_1^2-E_1E_2+E_2^2}{144E_1^3E_2^3}$ & $\frac{E_1^2-E_1E_2+E_2^2}{64E_1^3E_2^3}$  \\
     &   &       &            &           \\   \hline
\end{tabular}
\end{table}

where the function ${\rm S(k)}$ contains the powers of ${\rm k^0}$ or
${\rm ({\bf k} {\bf p})}$. The dimensional regularization was also used
for the calculation of the integrals (48). The corresponding results
are presented in the Table II where we wrote only the finite parts of the
obtained expressions.

Summing all contributions in (43), (44) after integration (48) we find that
the spin-independent part of the two-photon quasipotential has the
following structure: ${\rm (Z\alpha^2 {\bf p}^2 F(m_1,m_2)}$.
The contribution of this term to the energy spectrum can be obtained
by the use of the dimensional regularization and the equation (9)
\cite{CMY,PR}:
\begin{eqnarray}
{\rm <{\bf p}^2>=\psi_M(0)\int\frac{d{\bf p}}{(2\pi)^d}{\bf p}^2\psi({\bf p})=
b^2|\psi_M(0)|^2.}
\end{eqnarray}
So two-photon diagrams (a) and (b) Fig.1 yield the following correction
of order ${\rm (Z\alpha)^7}$ for the positronium spectrum in the hard
momentum region:
\begin{equation}
{\rm \Delta B_n(Ps)=-\frac{(Z\alpha)^7}{\pi n^5}m_1\frac{1129}{720}.}
\end{equation}
In the case of the muonium we can make additional expansion on the
powers ${\rm m_1/m_2}$. Preserving only linear terms on ${\rm m_1/m_2}$
we obtain:
\begin{equation}
{\rm \Delta B_n(Mu)=-\frac{(Z\alpha)^7}{\pi n^5}m_1\left(\frac{16}{5}+20\frac{m_1}{m_2}\right).}
\end{equation}
We took into account here only power corrections ${\rm (Z\alpha)^7}$.
The logarithmic terms of the kind ${\rm \ln(m_{1,2}/\lambda)}$
($\lambda$ is the parameter of the dimensional regularization)
were omitted systematically while they are present in the intermediate
expressions (48).

In the soft momentum region the general structure of the quasipotential
which gives the corrections of order ${\rm (Z\alpha)^7}$ is similar to (20).
In the leading order in ${\rm m_1/m_2}$ the muonium quasipotential has the
following form:
\begin{eqnarray}
{\rm V_{2\gamma}^{soft}({\bf p,
q})=\frac{i(Z\alpha)^2}{4\pi^2m_1^3m_2}\int
\frac{d^4k}{(k-p)^2(k-q)^2k_0^2}\Biggl\{\frac{(k^2-W^2)^4}{k_0^4}+}
\end{eqnarray}
\begin{displaymath}
{\rm +\frac{k^2-W^2)^3(k^2+2{\bf k}({\bf p}+{\bf q})-{\bf t}^2-W^2)}{k_0^4}-
\frac{3(k^2-W^2)^3}{k_0^2}+8W^2(k^2-W^2)-}
\end{displaymath}
\begin{displaymath}
{\rm -\frac{(k^2-W^2)^2(k^2+2{\bf k}({\bf p}+{\bf q})-{\bf
t}^2-W^2)}{k_0^2} -\frac{W^2(k^2-W^2)(k^2+2{\bf k}({\bf p}+{\bf
q})-{\bf t}^2-W^2)}{k_0^2}-}
\end{displaymath}
\begin{displaymath}
{\rm -\frac{W^2(k^2-W^2)^2}{k_0^2}+
\frac{4(2k_0^2-{\bf k}^2)(k^2-W^2)^2}{k_0^2}+\frac{2{\bf k}({\bf p}+{\bf q})
(k^2-W^2)^2}{k_0^2}\Biggr\}.}
\end{displaymath}
The residues in the photon poles determine the value of the integration
over ${\rm k_0}$ variable. The integration on ${\rm {\bf k}}$ can be
performed by standard methods by means of the dimensional regularization.
So in the soft approximation we obtain the following momentum
representation of the quasipotential (52):
\begin{eqnarray}
{\rm V^{soft}_{2\gamma}({\bf p},{\bf q})=\frac{(Z\alpha)^2}{2m_1^3m_2}
\Biggl\{\frac{13}{4}{\bf p}^2-\frac{W^2}{6}-\frac{(17{\bf p}^4-9W^4)}
{2t^2}+}
\end{eqnarray}
\begin{displaymath}
{\rm +\frac{8{\bf p}^6-4W^6+3{\bf p}^4(2{\bf q}^2+3W^2)+3{\bf
p}^2W^2({\bf q}^2-2W^2)} {3t^4}+}
\end{displaymath}
\begin{displaymath}
{\rm \frac{16W^8-29{\bf p}^8+16{\bf p}^6({\bf q}^2-2W^2)+3{\bf
p}^4(12W^4-{\bf q}^4)+ 4W^4{\bf p}^2(8W^2-9{\bf q}^2)}{12t^6}+}
\end{displaymath}
\begin{displaymath}
{\rm +\frac{(44{\bf p}^2-5W^2)}{8}\ln(t/W)+
\frac{14W^4-45{\bf p}^4+38{\bf p}^2{\bf q}^2+26W^2{\bf p}^2}{4t^2}\ln(t/W)+}
\end{displaymath}
\begin{displaymath}
{\rm +\frac{28W^6-8{\bf p}^6-53W^2{\bf
p}^4+W^2{\bf p}^2{\bf q}^2- 16W^4{\bf p}^2} {4t^4}\ln(t/W)+}
\end{displaymath}
\begin{displaymath}
{\rm +\frac{23{\bf p}^8-16W^8+16{\bf p}^6(2W^2-{\bf q}^2)+
{\bf p}^4(9{\bf q}^4-12W^4)+ 4W^4{\bf p}^2(3{\bf q}^2-8W^2)}
{4t^6}\ln(t/W)\biggr\}}
\end{displaymath}
The singularity of the different quasipotential terms (53) grows.
The averages of the large negative powers of r
${\rm <\frac{1}{r^3}>}$, ${\rm <\frac{1}{r^4}>}$, $\rm {<\frac{1}{r^5}>}$
arise when we calculate the energy spectrum in the coordinate representation
The transformation of such matrix elements was also done on the basis
of relations (24)-(26), (30), (A4)-(A7) by means of the computer program
FeynCalc for the system "Mathematica" \cite{Mertig,W}.
The contribution of order ${\rm (Z\alpha)^7}$ to the muonium (hydrogen atom)
S-states is determined by the expression:
\begin{displaymath}
{\rm \Delta B_n^{soft}=\frac{(Z\alpha)^7\mu^5}{m_1^3m_2\pi n^3}\Biggl\{
\frac{63565}{288}-
\frac{8675}{72}C-\frac{38}{3}\pi^2-\frac{2521}{72}\psi(n)+38\psi'(n)+}
\end{displaymath}
\begin{displaymath}
{\rm +\frac{1}{n}\left[\frac{605147}{9216}-\frac{148257}{1024}C-
\frac{360673}{6144}\psi(n)\right]
+\frac{1}{n^2}\left[\frac{3698203}{36864}+\frac{7157}{6144}C+\frac{10757}
{12288}\psi(n)\right]+}
\end{displaymath}
\begin{eqnarray}
{\rm
+\frac{1}{n^3}\left[-\frac{81}{64}+\frac{47}{16}C+\frac{47}{16}\psi(n)\right]\Biggr\},}
\end{eqnarray}
where ${\rm \psi(z)=d\ln\Gamma(z)/dz}$, C=0.5772156649... is the Euler constant.

\section{Discussion of the results.}

The calculation of the higher order perturbative corrections in
$\alpha$ in the energy spectra of hydrogen-like systems is very
urgent task at present due to the growth of the experimental
accuracy in the measurement of the energy levels in many simple
atomic systems. The complexity of such calculations will be
enhanced with the growth of the order $\alpha$ even for one loop
Feynman amplitudes. The most part of the computer programs
employed for the calculation of Feynman diagrams is working with
the mass shell particles. In the case of the bound systems the
Taylor series of the Feynman amplitudes should be taken for the
extraction of the necessary order contributions. In the present
work we considered how to use these expansions for the calculation
of the corrections ${\rm (Z\alpha)^6m_1^2/m_2}$ and ${\rm
(Z\alpha)^7}$. The dimensional regularization is the most
important tool for the use of Taylor expansions. It gives the
possibility to work simultaneously with the ultraviolet and
infrared divergencies and to obtain Fourier transforms of the
momentum space quasipotentials with high degree of the
singularity. Many terms of the quasipotential ${\rm V_{2\gamma}}$
obtained in this work in the coordinate space lead to the
divergent matrix elements in the case of S-states (see
Eq.(39,53)). So we have formulated the reqqurence relation (30)
in d-dimensional space which allows to take into consideration
correctly all coefficients in the divergent terms. Using (42) and
the results of our work \cite{MF} we can present the total
contribution of order ${\rm (Z\alpha)^6m_1^2/m_2}$ for arbitrary
principal quantum number n as follows (spin-dependent part of the
quasipotential was taken into account in our calculation):
\begin{equation}
\rm {\Delta B^{tot}_n=\left\{\frac{5}{2} \ln 2-\frac{151}{36}-\frac{3}{n}+
\frac{1}{2n^2}+\frac{547}{72n^3}-\frac{7}{2}(-1)^n(C+\Psi(n)-1)\right\}
\frac{m_1^2(Z\alpha)^6}{m_2n^3}.}
\end{equation}
Numerical value of the contribution (55) to the fine structure
interval ${\rm 2^3S_1\div 1^3S_1}$ of the muonium and hydrogen atom
is equal to 0.045 MHz and 5.116 KHz.
The methods for the calculation of Feynman amplitudes used in this work can be
applied for other higher order corrections. Numerical values for the
contributions of order ${\rm (Z\alpha)^7}$ (50), (51) and (54)
obtained in the hard and soft energy regions (5), (7) for the gross
structure interval ${\rm (2S\div 1S)}$ in muonium, hydrogen atom and
positronium are equal correspondingly 0.092 MHz, 0.130 MHz and 0.066 MHz.

\begin{acknowledgments}
We are grateful to A.S.Yelkhovsky for useful discussion of Ref. \cite{CMY}.
The work was performed under the financial support of the Russian Foundation
for Fundamental Research (grant 00-02-17771), the Program "Universities
of Russia - Fundamental Researches" (grant 990192) and the Ministry of Education
(grant EOO-3.3-45).
\end{acknowledgments}

\appendix

\section{Matrix elements of the potentials in the coordinate representation.}

To calculate the matrix elements ${\rm <\frac{1}{r^m}>}$,
${\rm<\frac{\ln r}{r^m}>}$ we must know the radial wave functions in
d-dimensional coordinate space and the reqqurence relation (30).
The radial wave function satisfies to the following equation:
\begin{equation}
{\rm R''+\frac{d-1}{\rho}R'+\left(-\frac{1}{4}+\frac{B}{\sqrt{A}}\frac{1}{\rho}-
\frac{l(l+d-2)}{\rho^2}\right)R=0,}
\end{equation}
where dimensionless variable ${\rm \rho=2r\sqrt{A}}$, ${\rm B=\mu Z\alpha}$,
${\rm A=-2\mu E}$. The normalized solution of this equation can be
obtained by the standard methods. The energy spectrum is determined by
(23) and corresponding radial S-state wave function has the form:
\begin{equation}
{\rm R_{n,l=0}(r)=\sqrt{\frac{2^dW^d\Gamma(n)}{(2n+d-3)\Gamma(n+d-2)}}e^{-Wr}\cdot
L_{n-1}^{d-2}(2Wr),~~~W=\frac{\mu Z\alpha}{n+\frac{d-3}{2}}.}
\end{equation}
where Laguerre polynomials
\begin{equation}
{\rm L_n^\lambda(z)=\frac{1}{n!}z^{-\lambda}e^z\frac{d^n}{dz^n}\left(e^{-z}z^{\lambda+n}\right).}
\end{equation}
In addition to the power potential matrix elements there is need to
consider the averages with the logarithmic function ${\rm \ln r}$:
${\rm <\frac{\ln r}{r^m}>}$. Using (30) we can obtain the following
matrix elements of such operators in d-dimensional space:
\begin{equation}
{\rm <\frac{\ln \rho}{\rho^3}>=-\frac{8\epsilon_n}{(d-1)(d-3)}<\frac{1}{\rho}>-
\frac{4(d^2-4d+5)}{(d-3)^2(d-1)^2}<\frac{1}{\rho^2}>+\frac{4}{(d-1)(d-3)}
<\frac{\ln \rho}{\rho^2}>},
\end{equation}
\begin{equation}
{\rm <\frac{\ln \rho}{\rho^4}>=-\frac{d^2-4d+24}{d^2(d-4)^2}<\frac{1}{\rho^3}>+
\frac{6}{d(d-4)}<\frac{\ln \rho}{\rho^3}>-}
\end{equation}
\begin{displaymath}
{\rm -\frac{2\epsilon_n(d^2-4d+8)}{(d-4)^2d^2} <\frac{1}{\rho^2}>+
\frac{4\epsilon_n}{d(d-4)}<\frac{\ln \rho}{\rho^2}>.}
\end{displaymath}
The averages ${\rm <\frac{\ln r}{r^2}>}$, ${\rm <\frac{\ln r}{r}>}$
are finite in d=3 space but we must find them by the use wave function
(A2)  taking into account the addenda $\sim\epsilon$=
${\rm \frac{d-3}{2}}$ in the limit ${\rm d\rightarrow 3}$:
\begin{equation}
{\rm
<\frac{\ln\rho}{\rho}>=\frac{\psi(n+d-2)}{(2n+d-3)},}
\end{equation}
\begin{equation}
{\rm <\frac{\ln\rho}{\rho^2}>=\frac{1}{(d-2)(2n+d-3)}\left[
\psi(d-1)+\psi(d-2)-\psi(n+d-3)\right].}
\end{equation}

To note also that the matrix elements of the operators ${\rm \frac{1}{r^2}}$
and ${\rm\frac{1}{r}}$ in d-dimensional space can be calculated by
means of the Hamiltonian operator
\begin{equation}
{\rm H=-\frac{1}{2}\left[\frac{\partial^2}{\partial
r^2}+\frac{(d-1)}{r} \frac{\partial}{\partial
r}\right]+\frac{1}{2r^2}\cdot l(l+d-2)-\frac{a}{r}.}
\end{equation}
Using (23) and (A8) we obtain:
\begin{equation}
{\rm<\frac{1}{r^2}>^d=\frac{2}{2l+d-2}<\frac{\partial H}{\partial l}>=
\frac{2}{(2l+d-2)}\frac{a^2}{(n+\frac{d-3}{2})^3},}
\end{equation}
\begin{equation}
{\rm <\frac{1}{r}>^d=\frac{a}{(n+\frac{d-3}{2})^2}.}
\end{equation}

\end{document}